\documentclass[a4paper]{article}

\usepackage{INTERSPEECH2020}

\title{Do face masks introduce bias in speech technologies? \\The case of automated scoring of speaking proficiency.}
\name{Anastassia Loukina, Keelan Evanini, Matthew Mulholland, Ian Blood, Klaus Zechner}
\address{Educational Testing Service, NJ, USA}
\email{aloukina@ets.org, kevanini@ets.org, mmulholland@ets.org,
       iblood@ets.org, kzechner@ets.org}

\usepackage{color, textcomp}
\def\FM{{\sc facemask}}

\begin{document}

\maketitle

\begin{abstract}
The COVID-19 pandemic has led to a dramatic increase in the use of face masks worldwide. Face coverings can affect both acoustic properties of the signal as well as speech patterns and have unintended effects if the person wearing the mask attempts to use speech processing technologies. In this paper we explore the impact of wearing face masks on the automated assessment of English language proficiency. We use a dataset from a large-scale speaking test for which test-takers were required to wear face masks during the test administration, and we compare it to a matched control sample of test-takers who took the same test before the mask requirements were put in place. We find that the two samples differ across a range of acoustic measures and also show a small but significant difference in speech patterns. However, these differences do not lead to differences in human or automated scores of English language proficiency. Several measures of bias showed no differences in scores between the two groups. 
\end{abstract}

\section{Introduction}

The spread of the COVID-19 pandemic has led to a dramatic increase in the use of face masks worldwide. However, the use of face coverings can lead to transmission loss and can modify the acoustic properties of the speech signal. They may also impact the speech patterns of the person wearing the mask. Both of these factors may in turn have unintended effects if the person wearing the mask attempts to use speech processing technologies \cite{Coniam2005,Mendel2008,Llamas2008, Saeidi2015,Saeidi2016}.

In this paper we explore the impact of wearing face masks on the automated assessment of English language proficiency. We consider an example of a large-scale English proficiency test which includes a speaking section and is usually taken at an official test center. With few exceptions, test-takers did not typically wear any face coverings when taking the test prior to the COVID-19 pandemic. However, starting in early 2020, local regulations in some areas have required test-takers to wear face masks. For instance, all test-takers taking the test in Hong Kong starting in late-February 2020 were required to wear surgical masks throughout the duration of the test, including the speaking section. This gave us a unique opportunity to evaluate the impact of face masks using data collected under real-life conditions. 

The study has two main goals. Our first goal is to explore the validity of using automated speech scoring systems in a situation when test-takers are wearing face masks. When automated systems are used as part of the scoring pipeline for assessments, it is important that they do not introduce bias into the final scores. Automated scoring systems are likely to have been trained on the data from test-takers who were not wearing face masks. If the models trained on this population do not generalize well to recordings from test-takers who are wearing masks, the scores assigned by the system would be unfair \cite{Zieky2016}. Our study is the first to explore the impact of face masks on the validity of automated speech scoring.

Our second goal is to contribute to our knowledge of the effect face masks might have on speech technologies in general. Previous studies on this topic have typically used small corpora collected under laboratory conditions. In our case, we have access to a unique dataset collected under real-life conditions from speakers who were wearing face masks while engaging in an authentic task (taking the speaking test). Furthermore, since this dataset was collected in the context of a standardized test, we also have data from a large population of test-takers who engaged with the same assessment tasks and under the same conditions except that they were not wearing face masks. This allows us to compare the data from test-takers wearing face masks to a matched control sample. We compare the two samples across a wide range of acoustic and phonetic measures to obtain a better understanding of how face masks may affect speech production and speech processing.

\section{Speech when wearing a face mask}

Previous research on the effect of surgical masks in speech studies has considered three possible types of effects: the effect on transmission loss and other signal properties, the effect on speech perception and intelligibility, and finally the effect on speech production when wearing a mask. Some of these studies were done in a clinical \cite{Mendel2008,Radonovich2010,Atcherson2017} or forensic context \cite{Saeidi2015,Saeidi2016}, although \cite{Coniam2005} also looked at the effect of face masks in the context of language proficiency assessment. 

In terms of acoustic effects, most fabrics have been shown to lead to transmission loss, especially at higher frequencies \cite{Nute1973}. These results have also been confirmed for surgical face masks by \cite{Llamas2008} and \cite{Saeidi2016} who found the largest differences in frequencies above 4.5 KHz. 

The limited evidence so far suggests that these changes to spectral quality have little effect on speech intelligibility. Paradoxically, \cite{Mendel2008} reported that speech perception scores for stimuli from the Connected Speech Test recorded by 1 speaker \emph{with} the mask present for listeners with both normal and impaired hearing were significantly better than their scores without the mask. However, given that this difference was less than 1\%, they did not consider it clinically significant. No effect of face mask for listeners with normal hearing was also reported by \cite{Atcherson2017} and \cite{Llamas2008}, who used a similar approach. In a slightly different set-up where 16 nurses were paired with each other for a series of face-to-face tests, \cite{Radonovich2010} also found that word intelligibility scores for subjects wearing a surgical mask were not significantly different from controls\footnote{In control condition the speaker's face was obscured by a piece of cardboard}.

Less is known about the effect of face masks on speech technologies. \cite{Saeidi2015,Saeidi2016} collected a corpus for forensic automatic speaker recognition which contained 1.5 hours of recordings from 8 speakers wearing several types of face coverings including surgical masks and a control condition. They found that surgical masks had little effect on the accuracy of closed-set automatic speaker identification: the accuracy of identifying a speaker wearing a mask using a speaker model trained without the mask ('no mask') was 95.1\%, very close to the 95.2\% accuracy in 'no mask' - 'no mask' condition \cite{Saeidi2016}. 

Finally, in the study that is most relevant to the context of this paper, \cite{Coniam2005} considered the effect face masks had on an English speaking proficiency assessment administered by in-person examiners in Hong Kong. In this study 186 students took a mock oral exam twice, with and without surgical masks, and then completed a short questionnaire. The tests were rated by trainee teachers. The results of the study showed that face masks had no significant effect on scores assigned by the human raters although for some areas, such as pronunciation, it is possible that a significant effect could have been detected on a larger sample. 

To our knowledge, there are no empirical studies of the effect the face masks might have on speaker behavior. The analysis of questionnaires in \cite{Coniam2005} provides some qualitative evidence in this direction: the test-takers felt that they performed at a lower level when wearing face masks and found it more difficult to understand their peers. The study also revealed some of the strategies used to compensate for face masks as reported by the speakers and raters: 92\% felt they spoke slower and 89\% felt they spoke louder. 12 out of 15 raters also observed that students might have been speaking louder and more slowly, or articulated more clearly. 

We use a large corpus of spoken responses recorded under real-life conditions to address the following research questions:  (1) What are the acoustic differences in responses recorded with and without face masks? (2) Are there any differences in speech patterns between the test takers who are wearing face masks and those who are not?  (3) Do face masks lead to a bias in performance of an automated scoring engine for spoken language proficiency?

\section{Data}

The data set considered in this paper includes responses from 597 test takers sampled from the data collected during operational administration of a large-scale language proficiency test in Hong Kong. The data set consists of two samples: \FM~and {\sc control}. 

Our \FM~sample consists of 297 test-takers who took an English language proficiency test in Hong Kong in late-February - March 2020, during the time when a requirement to wear surgical face masks was in place. Of these, 75\% reported Chinese as their native language,\footnote{The metadata we have available does not differentiate between different linguistic varieties.} 8\% reported Korean, with a further 10\% of speakers reporting English, Japanese, German or Indonesian. Other native languages were represented by less than 1\% of the test-takers. 50\% of the test-takers in this sample reported their gender as ``Female''. 

All test-takers took four components of the test: reading, listening, speaking and writing. The total duration of the test is about 3 hours. For the speaking section which takes about 20 minutes each test-taker provided responses to 4 questions that were designed to elicit spontaneous speech. For some questions test-takers were expected to use provided materials (e.g., a reading passage) as the basis for their response, while other questions were more general and elicited personal opinions or narratives. Depending on the question type, the speakers were given 45 seconds or 1 minute to complete their response. The audio was captured through a headset with a microphone in .ogg format. The final corpus consisted of 1,188 spoken responses. 

We also selected a matched {\sc control} sample from test-takers who took the test in Hong Kong in Fall 2019, before the requirement to wear masks took effect. We selected a stratified sample of 300 test-takers matched to the \FM~sample based on the proportion of speakers of different languages for the 6 languages that occurred most frequently in the \FM~sample. The test-takers in this sample took the test under the same conditions as the test-takers in the \FM~sample: they answered the same types of questions and their answers were recorded using the same equipment. The final corpus of control responses consisted of 1,200 spoken responses. 

The responses in both samples were scored using the same procedure: the audio recordings were sent to the distributed network of human raters who scored the responses on a 1-4 scale according to the scoring rubrics. The rating process is organized such that different responses from the same person are never scored by the same rater. In both samples, the responses from Hong Kong were scored together with responses from test centers in other countries. The raters were not aware of the country where the response was recorded. 

\section{Automated scoring engine}
 
All responses were also scored by SpeechRater\textsuperscript{\textregistered}, ETS's engine for scoring spoken responses to language proficiency tests \cite{Zechner2009a, Higgins2011a, Chen2018}.

All responses were first processed using an automated speech recognition system built using the Kaldi toolkit \cite{Povey2011}.
For the training of the acoustic model we used 800 hours of spoken responses of non-native spontaneous speech, covering over 100 native languages across almost 9,000 different speakers.  The DNN model was adapted to speakers with fMLLR and i-vectors using Kaldi's nnet2 DNN environment. The language model is a trigram model trained using the same dataset used for acoustic model training and is based on a vocabulary of around 30k lexical entries. The ASR training corpus was elicited using questions similar to the ones considered in this study. There was no overlap of speakers or questions between the ASR training corpus and the corpus used in this paper. We did not additionally adapt the ASR to the speakers or responses in this study. Detailed information about the ASR model building approach is  provided in \cite{Qian2016}.


The scoring model used to produce automated scores included 28 features. The scoring model was trained on a large sample of 500,000 responses from earlier administrations of the test worldwide. It was not adapted to the current sample. The features included in the model cover aspects of the delivery and language use, two dimensions of speaking proficiency that are considered by the human raters. Features related to delivery covered general fluency, pronunciation and prosody. Features related to language use covered vocabulary, grammar and some aspects of discourse structure. An additional module was used to flag atypical responses where an automated score is likely to be unreliable \cite{Higgins2011a, Yoon2018}.  See \cite{Chen2018} for a detailed description of the features and the filtering module. 

\section{Results}

\subsection{Face mask effect on human scores}

Even though the focus of this paper is the effect of face masks on speech technologies, it is conceivable that potential discomfort of wearing face masks would affect test-taker performance on other sections of the test (listening, reading, writing). Furthermore, there could be additional factors that would lead to score differences between the two samples in our study that are not directly related to wearing face masks. 

To address this we first used a mixed linear regression model to evaluate whether there was a significant difference between the two samples in scores for each section of the test.\footnote{All mixed-effects models reported in this paper were fit using version \texttt{lme4} package \cite{Bates2015}. Significance tests for parameter estimates were performed using \texttt{lmerTest} package \cite{Kuznetsova2017}.} The linear model has section score as the dependent variable, test taker as random factor and included sample and section as well as their interaction as fixed factors. We considered all four sections: speaking, listening, reading and writing. 

While there were differences in score distributions between different sections, the results showed that face masks had no effect on section scores with $p$ value varying from 0.36 to 0.97. In other words, there were no differences in means between the two samples, either in the human scores for the speaking section or in the scores for the other sections. 

We further considered whether there were differences in human scores at the level of individual speaking responses using response score as dependent variable, sample as fixed factor and test taker as random factor. The results were consistent with those obtained for section scores: the fact that some of the test-takers were wearing face masks during the speaking section had no effect on the average human scores assigned to their responses (2.95 for {\sc control} sample vs. 2.97 for \FM~sample, $p$=0.657).  

\subsection{Face mask effect on acoustic properties}

To examine the impact of wearing face masks on the acoustic properties of test-taker responses, we extracted a range of acoustic features using the openSMILE toolkit \cite{Eyben2010}.  We extracted the 88 features in the extended GeMAPS set, which included measurements related to frequency, energy/amplitude and spectral characteristics. 

Unpaired, two-sample t-tests on the distributions of the feature values for the \FM~and {\sc control} samples indicate that 38 out of the 88 openSMILE features have significant differences between the two samples (using Bonferroni correction to adjust the significance level for multiple comparisons).\footnote{Two of the control responses were excluded from the analysis, resulting in a total of 1198 control responses, since openSMILE features could not be extracted for them.} A few examples of features that have distinct distributions include mean slope of the power spectrum between the 0Hz and 500Hz bands and mean bandwidth value for F2. In order to investigate whether responses from the two samples can be distinguished using these features, classification experiments were conducted using the SKLL machine learning toolkit\footnote{https://github.com/EducationalTestingService/skll}.  Specifically, 10-fold cross-validation (with no overlap of responses from the same test-taker across folds) was used to train and evaluate a range of models; the best performing model (\texttt{GradientBoostingClassifier}) achieved an F1-score of 0.786 and accuracy of 0.783.

\subsection{Face mask effect on speech patterns}

\begin{figure*}[!t]
\includegraphics[scale=0.34]{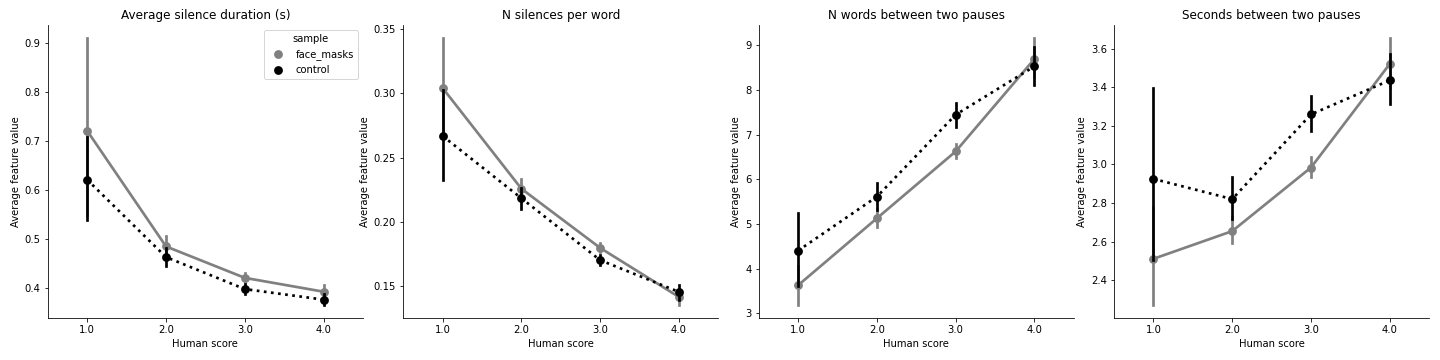}
  \caption{Average values for different speech properties for the two groups of test-takers conditioned on the human score.}
  \label{fig:dff_feature}
\end{figure*}

We next considered whether face masks had any effect on test-taker speech patterns. Using ASR hypotheses and timestamps we computed 8 different features designed to capture whether test-takers wearing masks made more pauses, spoke more slowly or showed different patterns of disfluencies. Since all these features depend on speaker proficiency, for each feature we first fit a baseline linear model with the feature value as dependent variable and human score as predictor. We then added the sample as a second independent variable and computed the difference in adjusted $R^2$ between the two models. The results are shown in Table \ref{tab:patterns} and Figure \ref{fig:dff_feature}. 

Four out of eight properties we considered showed significant differences: average duration of silences and number of silences per word as well as the duration of chunks between pauses, whether measured in words or seconds. In other words, speakers wearing masks spoke with about the same articulation rate as those not wearing masks, but paused slightly more often. However, the difference between the two samples was very small: face masks only explained less than 1\% variance in feature values (after controlling for proficiency). In absolute terms, wearing masks reduced the duration of chunks between two pauses by 0.6 words or 0.2 seconds. 

\begin{table}[!ht]
\caption{The effect of speech masks on various speech patterns. The table shows the additional variance in feature value explained by {\sc face mask} after controlling for human score, standardized coefficient and whether the result was statistically significant after Bonferroni correction for multiple comparison.}
\label{tab:patterns}
\centering
\begin{tabular}{llrrl}
 Feature &  $R^2$ diff &  Coef. & sig \\
\midrule
Average duration of silences &    0.006 &            0.021 &   * \\
N silences per word &    0.003 &            0.007 &   * \\
Total N silences &    0.001 &            0.382 &  ns \\
Seconds between two pauses &    0.009 &           -0.188 &   * \\
N words between two pauses &    0.008 &           -0.567 &   * \\
Total N words &    0.002 &           -2.278 &  ns \\
Words per second of speech &    0.000 &           -0.014 &  ns \\
Number of disfluencies &    0.002 &           -0.563 &  ns \\
\bottomrule
\end{tabular}
\end{table}

\subsection{Facemask effect on ASR performance}

To further explore the effect face masks might have on ASR performance, we selected a sample of 55 responses (28 \FM~responses and 27 {\sc control} responses) and had them transcribed by a group of three transcribers. Of these, 16 responses were triple-transcribed. We then computed ASR word error rate (WER) for each group. A Mann Whitney U test showed that there was no difference in WER between the two samples (29.6\% for {\sc control} sample vs. 27.7\% for \FM~sample, $p$=0.22). The WER between two humans varied between 19.7\% and 27.6\% depending on the pair of transcribers. Surprisingly, human-human WER seemed to be lower for \FM~sample (27.6\% vs. 20.7\%, linear regression model after controlling for transcriber pair $p$=0.04). We note however that the sample size is very small.

\subsection{Face mask effect on automated scores}

Finally, we considered whether the small differences we observed between the two samples had any effect on automated scores computed for the responses. First of all, we looked into whether more responses from test-takers with face masks have been flagged as non-scorable by the SpeechRater filters. We found this not to be the case: only 1 response in the \FM~sample was flagged as non-scorable in comparison to 3 responses in the {\sc control} sample. In addition, 11 responses in the {\sc control} sample and 9 responses in the \FM~sample were not sent for automated scoring. 

We first consider the effect of face masks on the outcome fairness: that is, whether the automated scores assigned to responses are affected by whether the test-taker is wearing a mask. We considered overall accuracy of scores as well as overall score differences, that is, whether the automated scores are consistently different from human scores for members of a certain group \cite{Williamson2012,Loukina2019}. We used RSMTool \cite{Madnani2017} to compute all metrics.

\begin{table}[!h]
\caption{Accuracy of automated scores for responses from test-takers wearing masks and the control sample. The table shows mean system score, Pearson's $r$, and root mean squared error (RMSE).}
\label{tab:evaluation}
\centering
\begin{tabular}{lrrrr}
\toprule
Sample  &       N & Sys mean & $r$ &  RMSE \\
\midrule
{\sc control}    & 1186 &   2.79    &    0.56 &          0.58  \\
{\sc facemask} & 1178 &  2.80       &  0.58 &         0.55 \\
\bottomrule
\end{tabular}
\end{table}

Table \ref{tab:evaluation} shows the results. The accuracy of automated scores was the same on both samples: thus, for example, Pearson's $r$ between human and system scores was 0.56 for the {\sc control} sample and 0.58 for the \FM~sample. Test for significance of difference between two independent correlations using the Fisher r-to-z transform showed that the difference was not significant ($p$=0.44). There also was no significant difference in root mean squared error (RMSE) between the two samples ($p$=0.06) (``overall score accuracy'' in \cite{Loukina2019}). Finally, there was no difference in the actual scores:  the differences between standardized means for both samples were below the 0.01 threshold and there was no significant difference in the absolute error between machine and human score ($p$=0.678) (``overall score difference'' in \cite{Loukina2019}).

We also considered process fairness, that is whether the automated scoring engine assigns different scores to test-takers from different samples despite them having the same proficiency (``conditional score accuracy'' in \cite{Loukina2019}). To do this we looked into how much additional variance in score error is explained by sample membership after controlling for human score. The results once again showed no difference between test-takers in the two samples ($p$=0.853). 
    
\subsection{Discussion and conclusion}

In this paper we used a large corpus of responses to language proficiency assessment to evaluate the effect face masks may have on different aspects of speech processing. 

Face masks had an effect on various acoustic properties of the signal: our classifier experiments showed that it is possible to predict with almost 80\% accuracy whether a test-taker is wearing a mask or not based on low-level OpenSmile features. We also found that face masks led to small but significant differences in speech patterns: test-takers wearing masks tended to pause a bit more often than the control sample. The difference was very small: 0.6 words or 0.2 seconds. 

However, these differences in acoustics and speech patterns did not have a further effect on the performance of automated speech recognition or the automated scoring engine. We found no difference in ASR WER. Automated scores also were not affected by whether the test-takers were wearing masks or not. Finally, we found no difference in proficiency scores assigned by human raters. These results are very encouraging: as of the time of writing this paper, the use of face masks remains widespread. It is important to know that they can continue to be used in the context of language proficiency assessment without having negative impacts on test scores. 

To our knowledge, this is the first study that has explored the use of face masks using a large number of speakers and a corpus obtained in real-life conditions. Our results are consistent with previous research conducted with a smaller number of subjects in controlled laboratory conditions, which showed no effect of surgical face masks on speech intelligibility by humans. While these results may appear counter-intuitive given the widespread perception that face masks degrade intelligibility, we note that in face-to-face communications the use of face coverings also results in the reduction in visual information available to listeners. Multiple studies since the 1930s have shown that being able to see a speaker’s mouth movements greatly enhances intelligibility, especially where speech is presented in noise \cite{Llamas2008}. Both this paper and previous studies (except \cite{Coniam2005}) considered situations where the listeners and the automated systems had access only to the auditory stimulus collected with or without face mask and without additional visual information. 

To conclude, our analysis of speech patterns and acoustic features in combination with previously published results suggests that face masks are unlikely to have a substantial detrimental effect on speech technologies. Yet the differences we observed for low-level acoustic features suggest that some types of technologies and applications may be more affected than others. 



\begin{thebibliography}{10}
\providecommand{\url}[1]{#1}
\csname url@samestyle\endcsname
\providecommand{\newblock}{\relax}
\providecommand{\bibinfo}[2]{#2}
\providecommand{\BIBentrySTDinterwordspacing}{\spaceskip=0pt\relax}
\providecommand{\BIBentryALTinterwordstretchfactor}{4}
\providecommand{\BIBentryALTinterwordspacing}{\spaceskip=\fontdimen2\font plus
\BIBentryALTinterwordstretchfactor\fontdimen3\font minus
  \fontdimen4\font\relax}
\providecommand{\BIBforeignlanguage}[2]{{%
\expandafter\ifx\csname l@#1\endcsname\relax
\typeout{** WARNING: IEEEtran.bst: No hyphenation pattern has been}%
\typeout{** loaded for the language `#1'. Using the pattern for}%
\typeout{** the default language instead.}%
\else
\language=\csname l@#1\endcsname
\fi
#2}}
\providecommand{\BIBdecl}{\relax}
\BIBdecl

\bibitem{Coniam2005}
D.~Coniam, ``The impact of wearing a face mask in a high-stakes oral
  examination: An exploratory post-{SARS} study in {H}ong {K}ong,''
  \emph{Language Assessment Quarterly}, vol.~2, no.~4, pp. 235--261, 2005.

\bibitem{Mendel2008}
L.~L. Mendel, J.~A. Gardino, and S.~R. Atcherson, ``{Speech understanding using
  surgical masks: A problem in health care?}'' \emph{Journal of the American
  Academy of Audiology}, vol.~19, no.~9, pp. 686--695, 2008.

\bibitem{Llamas2008}
C.~Llamas, P.~Harrison, D.~Donnelly, and D.~Watt, ``Effects of different types
  of face coverings on speech acoustics and intelligibility,'' \emph{York
  Papers of Linguistics Series 2}, no.~9, pp. 80--104, 2008.

\bibitem{Saeidi2015}
R.~Saeidi, T.~Niemi, H.~Karppelin, J.~Pohjalainen, T.~Kinnunen, and P.~Alku,
  ``{Speaker recognition for speech under face cover},'' \emph{Proceedings of
  the Annual Conference of the International Speech Communication Association,
  INTERSPEECH}, vol. 2015-Janua, pp. 1012--1016, 2015.

\bibitem{Saeidi2016}
R.~Saeidi, I.~Huhtakallio, and P.~Alku, ``{Analysis of face mask effect on
  speaker recognition},'' \emph{Proceedings of the Annual Conference of the
  International Speech Communication Association, INTERSPEECH}, vol.
  08-12-Sept, pp. 1800--1804, 2016.

\bibitem{Zieky2016}
M.~J. Zieky, ``Fairness in test design and development,'' in \emph{Fairness in
  Educational Assessment and Measurement}, N.~J. Dorans and L.~L. Cook,
  Eds.\hskip 1em plus 0.5em minus 0.4em\relax Routledge, 2016, pp. 9--32.

\bibitem{Radonovich2010}
L.~J. Radonovich, R.~Yanke, J.~Cheng, and B.~Bender, ``{Diminished speech
  intelligibility associated with certain types of respirators worn by
  healthcare workers},'' \emph{Journal of Occupational and Environmental
  Hygiene}, vol.~7, no.~1, pp. 63--70, 2010.

\bibitem{Atcherson2017}
S.~R. Atcherson, L.~L. Mendel, W.~J. Baltimore, C.~Patro, S.~Lee, M.~Pousson,
  and M.~J. Spann, ``The effect of conventional and transparent surgical masks
  on speech understanding in individuals with and without hearing loss,''
  \emph{Journal of the American Academy of Audiology}, vol.~28, no.~1, 2017.

\bibitem{Nute1973}
M.~E. Nute and K.~Slater, ``The effect of fabric parameters on
  sound-transmission loss,'' \emph{The Journal of The Textile Institute},
  vol.~64, no.~11, pp. 652--658, 1973.

\bibitem{Zechner2009a}
K.~Zechner, D.~Higgins, X.~Xi, and D.~M. Williamson, ``{Automatic scoring of
  non-native spontaneous speech in tests of spoken English},'' \emph{Speech
  Communication}, vol.~51, no.~10, pp. 883--895, 2009.

\bibitem{Higgins2011a}
D.~Higgins, X.~Xi, K.~Zechner, and D.~Williamson, ``{A three-stage approach to
  the automated scoring of spontaneous spoken responses},'' \emph{Computer
  Speech {\&} Language}, vol.~25, no.~2, pp. 282--306, 2011.

\bibitem{Chen2018}
\BIBentryALTinterwordspacing
L.~Chen, K.~Zechner, S.-Y. Yoon, K.~Evanini, X.~Wang, A.~Loukina, J.~Tao,
  L.~Davis, C.~M. Lee, M.~Ma, R.~Mundkowsky, C.~Lu, C.~W. Leong, and
  B.~Gyawali, ``Automated scoring of nonnative speech using the
  SpeechRater. 5.0 engine,'' \emph{ETS Research Report
  Series}, vol. 2018, no.~1, pp. 1--31, dec 2018. [Online]. Available:
  \url{http://doi.wiley.com/10.1002/ets2.12198}
\BIBentrySTDinterwordspacing

\bibitem{Povey2011}
D.~Povey, A.~Ghoshal, G.~Boulianne, L.~Burget, O.~Glembek, N.~Goel,
  M.~Hannemann, P.~Motlicek, Y.~Qian, P.~Schwarz, J.~Silovsky, G.~Stemmer, and
  K.~Vesely, ``The {Kaldi} speech recognition toolkit,'' in \emph{Proceedings
  of the Workshop on Automatic Speech Recognition and Understanding}, 2011.

\bibitem{Qian2016}
Y.~Qian, X.~Wang, K.~Evanini, and D.~Suendermann-Oeft, ``{Self-adaptive DNN for
  improving spoken language proficiency assessment},'' \emph{Proceedings of the
  Annual Conference of the International Speech Communication Association,
  INTERSPEECH}, vol. 08-12-September-2016, pp. 3122--3126, 2016.

\bibitem{Yoon2018}
\BIBentryALTinterwordspacing
S.-Y. Yoon, A.~Cahill, A.~Loukina, K.~Zechner, B.~Riordan, and N.~Madnani,
  ``Atypical inputs in educational applications,'' in \emph{Proceedings of the
  2018 Conference of the North American Chapter of the Association for
  Computational Linguistics: Human Language Technologies, Volume 3 (Industry
  Papers)}.\hskip 1em plus 0.5em minus 0.4em\relax Stroudsburg, PA, USA:
  Association for Computational Linguistics, 2018, pp. 60--67. [Online].
  Available: \url{http://aclweb.org/anthology/N18-3008}
\BIBentrySTDinterwordspacing

\bibitem{Bates2015}
D.~Bates, M.~M{\"{a}}chler, B.~Bolker, and S.~Walker, ``Fitting linear
  mixed-effects models using lme4,'' \emph{Journal of Statistical Software},
  vol.~67, no.~1, pp. 1--48, 2015.

\bibitem{Kuznetsova2017}
A.~Kuznetsova, P.~B. Brockhoff, and R.~H.~B. Christensen, ``{lmerTest} package:
  Tests in linear mixed effects models,'' \emph{Journal of Statistical
  Software}, vol.~82, no.~13, pp. 1--26, 2017.

\bibitem{Eyben2010}
\BIBentryALTinterwordspacing
F.~Eyben, M.~W\"{o}llmer, and B.~Schuller, ``{openSMILE} – the {Munich}
  versatile and fast open-source audio feature extractor,'' in
  \emph{Proceedings of the 18th ACM international conference on
  Multimedia}.\hskip 1em plus 0.5em minus 0.4em\relax Firenze, Italy:
  Association for Computing Machinery, 2010, pp. 1459--1462. [Online].
  Available: \url{https://dl.acm.org/doi/pdf/10.1145/1873951.1874246}
\BIBentrySTDinterwordspacing

\bibitem{Williamson2012}
D.~M. Williamson, X.~Xi, and F.~J. Breyer, ``A framework for evaluation and use
  of automated scoring,'' \emph{Educational Measurement: Issues and Practice},
  vol.~31, no.~1, pp. 2--13, 2012.

\bibitem{Loukina2019}
\BIBentryALTinterwordspacing
A.~Loukina, N.~Madnani, and K.~Zechner, ``{The many dimensions of algorithmic
  fairness in educational applications},'' in \emph{Proceedings of the
  Fourteenth Workshop on Innovative Use of NLP for Building Educational
  Applications}.\hskip 1em plus 0.5em minus 0.4em\relax Florence, Italy:
  Association for Computational Linguistics, 2019, pp. 1--10. [Online].
  Available: \url{https://www.aclweb.org/anthology/W19-4401}
\BIBentrySTDinterwordspacing

\bibitem{Madnani2017}
N.~Madnani, A.~Loukina, A.~von Davier, J.~Burstein, and A.~Cahill, ``Building
  better open-source tools to support fairness in automated scoring,'' in
  \emph{Proceedings of the EACL Workshop on Ethics in Natural Language
  Processing}, 2017, pp. 41--52.

\end{thebibliography}
\end{document}